\documentclass[aps,prl,twocolumn,groupedaddress,showpacs,floatfix,superscriptaddress,longbibliography]{revtex4-2}
\usepackage[plainpages=false,pdfpagelabels,colorlinks=true,linkcolor=red,urlcolor=blue,citecolor=blue,pdftitle={Title},pdfauthor={},pdfdisplaydoctitle=true,pdfduplex=DuplexFlipLongEdge]{hyperref}
\usepackage{epsfig}
\usepackage{graphicx}
\usepackage[utf8]{inputenc}
\usepackage{amsmath,amssymb}
\usepackage{physics}
\usepackage[dvipsnames]{xcolor}

\usepackage[normalem]{ulem}

\begin{document}

\title{
Site-selective correlations in interacting ``flat-band'' quasicrystals}

\author{Yuxi Zhang}
\affiliation{Department of Physics, University of Illinois Urbana-Champaign, Urbana, IL, USA}
\author{Richard T. Scalettar}
\affiliation{Department of Physics, University of California, Davis, CA, USA}
\author{Rafael M. Fernandes}
\affiliation{Department of Physics, University of Illinois Urbana-Champaign, Urbana, IL, USA}
\date{\today}

\begin{abstract}
Model lattices such as the kagome and Lieb lattices have been widely investigated to elucidate the properties of interacting flat-band systems. While a quasicrystal does not have proper bands, the non-interacting density of states of several of them displays the typical signature of a flat band pinned at the Fermi level: a delta-function zero-energy peak. Here, we employ quantum Monte Carlo simulations to determine the effect of onsite repulsion on these quasicrystals. While global properties such as the antiferromagnetic structure factor and the specific heat behave similarly as in the case of periodic lattices undergoing a Mott transition, the behavior of the local density of states depends on the coordination number of the site. In particular, sites with the smallest coordination number, which give the dominant spectral-weight contribution to the zero-energy peak, are the ones most strongly impacted by the interaction. Besides establishing site-selective correlations in quasicrystals, our work also points to the importance of the real-space structure of flat bands in interacting systems.
\end{abstract}
\maketitle

Flat-band systems provide a unique framework to expand our conceptual understanding of strongly-correlated electronic systems. Because of the vanishingly small bandwidth, the electron-electron interaction has a strong impact on the electronic spectrum, even if it is small in absolute value. For this reason, flat bands have been theoretically and experimentally investigated in diverse settings  \cite{Regnault2022,Neves2024}, from moir\'e systems tuned to special twist-angle values \cite{cao18co,cao18un,Bistritzer2011} to geometrically-frustrated lattices such as kagome \cite{meier20,kang20,Kang2020dirac,Ye2024} and pyrochlore \cite{Huang2024}.

A system that has been less explored in this context are quasi-periodic crystals \cite{shechtman84,levine84}. Since proper bands cannot be defined, the electronic spectrum of quasicrystals is assessed via the density of states (DOS). In a periodic lattice, a flat band is manifested as a delta-function peak in the DOS, since the integrated DOS is discontinuous. Interestingly, in several quasicrystals, including the Penrose tilings known as kite-and-dart (or P2) and rhombus (or P3), a delta-function peak pinned at zero energy is found in the DOS  \cite{kohmoto86,Arai1988,Rieth1995}. These peaks are the manifestation of a macroscopic number of zero-energy states, corresponding to approximately $10\%$ of all states in the Penrose tiling. Despite the similarities between the zero-energy peaks of the P2 and P3 tiling, their microscopic origins and real-space distributions are quite different  \cite{Day20}. 

\begin{figure*}
\centering
\includegraphics[width=2\columnwidth]{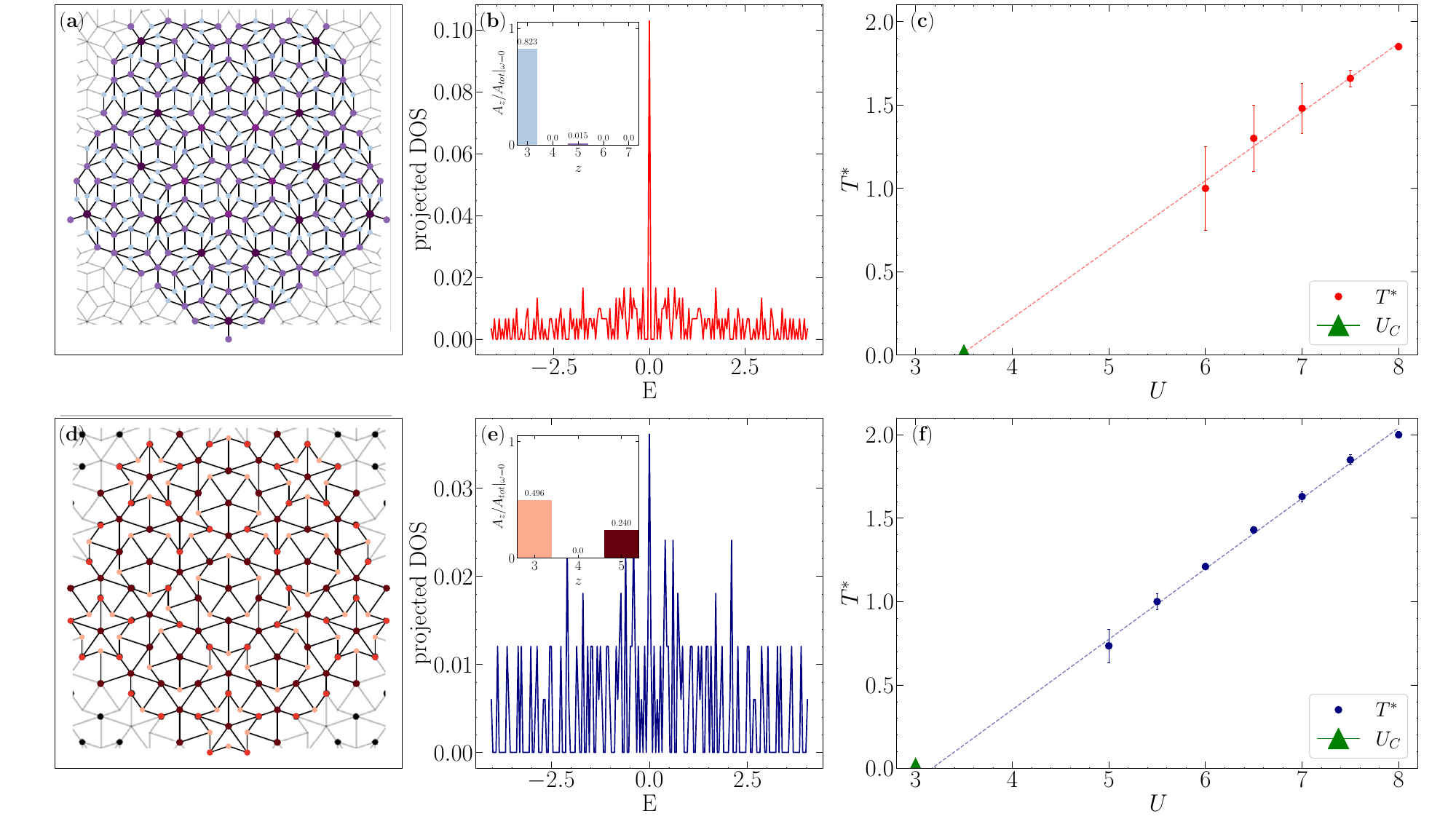}
\caption{(a) Penrose rhombus tiling; sites are colored according to their coordination number $3 \leq z \leq 7$. (b)
Density of states for the $N=301$ tiling studied here. Inset: Zero-energy spectral-weight ratio $A_z/A_{\mathrm{tot}} |_{\omega=0}$, highlighting the dominant contribution from the $z=3$ sites.
(c) Crossing temperature  $T^{*}$ of the DOS proxy curves, shown in Fig.~ \ref{fig:proxy}, as a function of $U$. The dashed line is a linear interpolation. Only non-zero $T^*$ values (red circles) that could be unambiguously resolved are shown. $U_c$ (green triangle) is the critical $U$ value extracted from the specific heat analysis. (d)-(f): Same as (a)-(c), but for the Penrose kite-and-dart tiling, where $3 \leq z \leq 5$ and $N=166$. In all figures of this paper, we set $t=1$.}
\label{fig:model}
\end{figure*}

Theoretical interest in interacting quasicrystals has surged in recent years \cite{Andrade2015,takemori15,Chalker2015,Arita2017,koga17,araujo19,Parameswaran2020,Tohyama2020,Yang2020,Arita2020,Gautier2021,Profe2021,sakai22Hyperuniform,sakai22,Gottlob2023,Keskiner2023,Yang2023,Yang2024,Gali2024,Hecker2024,Liu2024}, partly motivated by the experimental observation of correlated phenomena such as superconductivity  \cite{kamiya18,uri23,tokumoto24}, magnetism  \cite{Tamura2010,Ishikawa2018,Das23}, and quantum criticality \cite{deguchi12,Matsukawa2016,Khansili2024}. Moreover, quasicrystalline patterns have been engineered in diverse settings such as optical lattices \cite{jagannathan14,corcovilos19,yu20}, photonic lattices \cite{kohmoto87,vardeny13}, moir\'e superlattices \cite{uri23,tokumoto24,Liu2024field}, and synthetic lattices \cite{collins17} thus significantly expanding the opportunities to tune, probe, and control quasicrystals. Despite such an intense activity, a systematic investigation of the role of ``flat bands'' in interacting quasicrystals remains little explored.

In this paper, we elucidate the fate of ``flat-band'' quasicrystals in the presence of strong electron-electron interactions. In particular, we use the exact and unbiased determinantal quantum Monte Carlo (DQMC) method \cite{Blankenbecler81,scalettar86,white88} to solve the Hubbard model on both the rhombus and the kite-and-dart Penrose tiling. We find that global quantities such as the specific heat and the antiferromagnetic structure factor display similar signatures of a Mott transition as those observed in periodic lattices \cite{paiva05,varney09}, indicative of a Mott transition at $U^*/t \approx 3.5$  (see also  \cite{takemori15,koga17}). Here, $U$ is the onsite Hubbard interaction and $t$ is the nearest-neighbor hopping parameter. 

Motivated by the fact that these quasicrystals have sites with different coordination numbers $3 \leq z \leq 7$ , we also probe local properties by computing a proxy for the zero-energy local density of states (LDOS). Previous real-space dynamical mean-field theory (DMFT) calculations of the Hubbard model on a Penrose quasicrystal identified site-dependent double occupancy and renormalization factors whose origin remains unsettled  \cite{takemori15}. Our main result is the observation of site-selective correlations in the strong-coupling regime of $U > U^*$. While at high temperatures the LDOS proxy is the largest at the sites with the smallest coordination $z$, this trend reverses below a characteristic temperature $T^*$, where the sites with the smallest $z$ value display the smallest LDOS proxy values. Interestingly,  $T^*$ extrapolates to zero close to the characteristic interaction strength $U^*$ where the global Mott transition is estimated. By analyzing the frequency-dependent LDOS at each site via analytical continuation, we confirm that different sites are impacted in distinct ways by interactions, with larger gaps emerging for sites with smaller $z$. 
Interestingly, the sites that are the most strongly impacted by correlations are also those that contribute the most spectral weight to the non-interacting zero-energy peak of the global DOS, revealing a subtle interplay between interactions and flat bands. 

Our starting point is the half-filled Hubbard model \cite{,Arovas22,Gull2022}:
\begin{align}
\mathcal{\hat H} = & - \sum_{\langle \mathbf{i}, \mathbf{j}
  \rangle, \sigma} \big(t
\, \hat c^{\dagger}_{\mathbf{i} \sigma} \hat
c^{\phantom{\dagger}}_{\mathbf{j} \sigma} + {\rm H.c.} \big) \nonumber \\
+ & U \sum_{\mathbf{i}} \left( \hat
n_{\mathbf{i} , \uparrow} - \frac{1}{2}\right) \left( \hat
n_{\mathbf{i} , \downarrow} - \frac{1}{2}\right) \,\, ,
\label{eq:Hamiltonian}
\end{align}
where $\hat c^{\phantom{\dagger}}_{\mathbf{i}}$ is the annihilation operator for an electron on site $\mathbf{i}$ and spin $\sigma$, and $\hat n_\mathbf{i}=\hat c^{\dagger}_{\mathbf{i}} \hat c^{\phantom{\dagger}}_{\mathbf{i}}$ is the number operator. Hereafter, we set $t=1$.
The two types of Penrose quasicrystals studied here are shown in Figs.~\ref{fig:model}(a) and (d), corresponding to the rhombus and kite-and-dart tiling, respectively, both of which are bipartite. In the figure, the sites are colored according to their coordination number $z$ (see insets of panels (b) and (e)), which assume values in the range $3 \leq z \leq 7$ for the rhombus tiling and  $3 \leq z \leq 5$ for the kite-and-dart tiling. The sites marked in Figs.~\ref{fig:model}(a) and (d) form the finite-size aperiodic lattices with open boundary conditions used in our DQMC simulations, with $N=301$ sites (rhombus) and $N=166$ sites (kite-and-dart), constructed to explicitly preserve the five-fold symmetry of the Penrose geometry.

\begin{figure*}[htbp!]
\centering
\includegraphics[width=2\columnwidth]{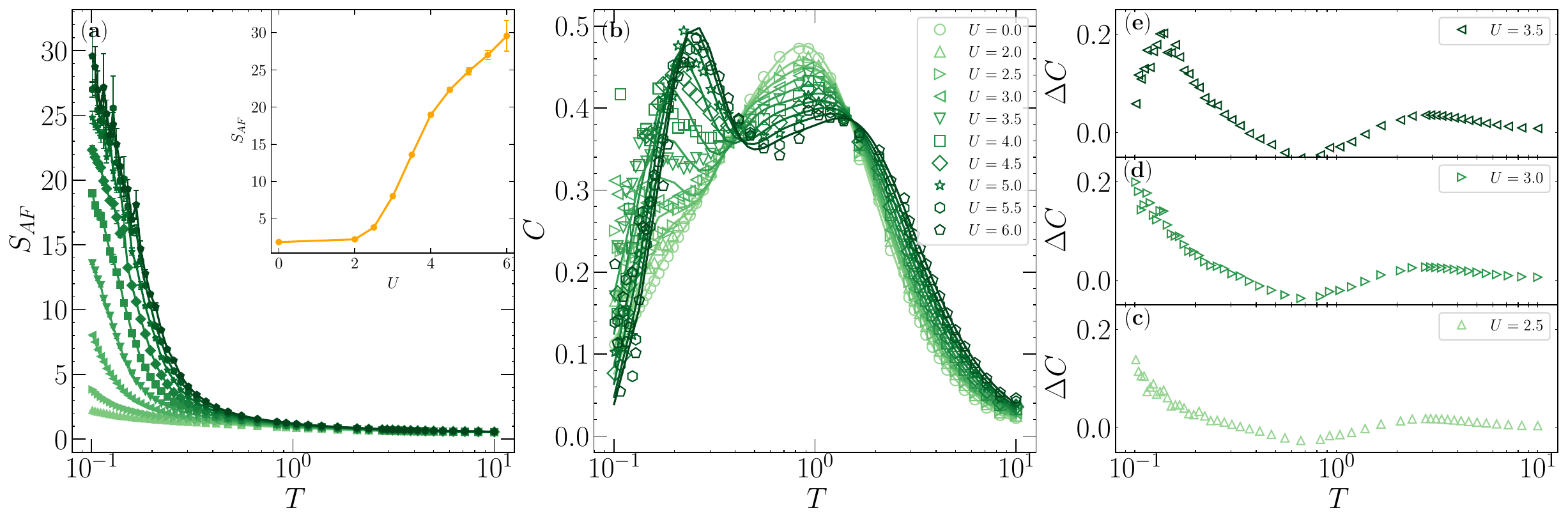}
\caption{Antiferromagnetic structure factor $S_{AF}$ (a) and specific heat $C$ (b) as functions of temperature $T$ for the rhombus tiling. The colors indicate different $U$ values, which increase as the color scale change from light to dark; the inset in (a) shows $S_{AF}$ at low temperature $T=0.1$.
(c)-(e) Specific heat difference with respect to the non-interacting case, $\Delta C \equiv C(U)-C(U=0)$. A low-temperature peak emerges at $U_c = 3.5$.}
\label{fig:mott}
\end{figure*}

The non-interacting  ($U=0$) electronic spectra of both 
tilings are characterized by delta-function zero-energy peaks in the DOS in the thermodynamic limit \cite{kohmoto86,Arai1988,Day20}. Figs.~\ref{fig:model} (b) and (e) demonstrate that the DOS of our finite-size tilings display these zero-energy peaks. Noticeably, not all sites contribute equally to the zero-energy peak.
When projected onto the coordination number $z$, the zero-energy DOS is found to be dominated by the sites with
$z=3$ 
in both geometries, as shown by the $z$-resolved relative spectral function calculated at zero-energy, $A_z/A_{\mathrm{tot}}|_{E=0}$ [insets of Figs.~\ref{fig:model} (b), (e)]. The sum is smaller than $1$ because the boundary sites are excluded due to the open boundary conditions used here. 

To investigate the impact of $U$, we solve Eq. (\ref{eq:Hamiltonian}) via DQMC, an exact and unbiased method that does not have the sign-problem since the model (\ref{eq:Hamiltonian}) is particle-hole symmetric (additional details, including the parameters used, are found in the Supplemental Material, SM~ \cite{sm}). 
Hereafter, we focus on the rhombus tiling results, leaving the 
kite-and-dart tiling results to the SM.

We start by analyzing the quasicrystal's global properties.
Fig. ~\ref{fig:mott} (a) displays the antiferromagnetic (AF) structure factor $S_{AF}$ as a function of temperature $T$ for multiple 
interaction strength values. While for small $U$ the system shows no sign of AF order down to the lowest temperatures probed, magnetic correlations are strongly enhanced for $U \gtrsim 3.0$ (see inset), indicative of an AF-Mott transition. Such a transition is also observed by RDMFT calculations, albeit at larger $U$ values   \cite{takemori15,sakai22}. This transition is further confirmed by analyzing the specific heat $C$, Fig.~\ref{fig:mott} (b). 
For $U \leq 3.5$, $C(T)$ shows a broad peak around $T\approx 1.0$, similar to the non-interacting system. However, for $U \geqslant 3.5$, $C(T)$ develops a sharper second peak at lower temperatures, as highlighted in Figs.~\ref{fig:mott} (c)-(e), where  $\Delta C\equiv C(U)-C(U=0)$ is plotted near $U=3.5$. 
A second specific-heat peak is also seen in DMFT and DQMC solutions of the Hubbard model in periodic lattices \cite{Georges1993,Vollhardt97,Vollhardt99,paiva05}, where it is attributed to AF correlations. 
Thus, we associate $U^* \approx 3.5$ to a putative AF-Mott transition.

\begin{figure}
\centering
\includegraphics[width=1\columnwidth]{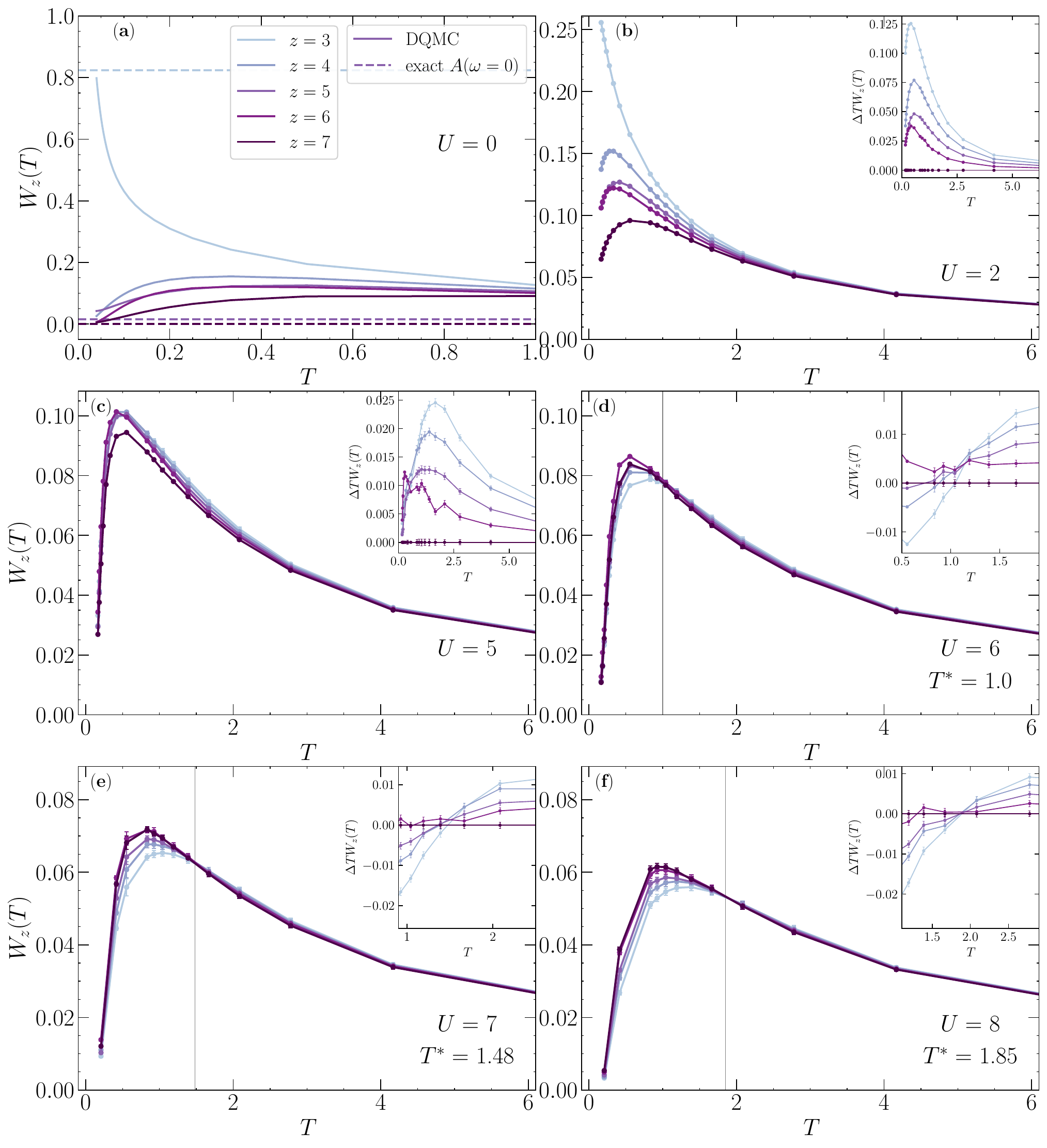}
\caption{Proxy $W_{z}(T)$ for the average local density of states at a site with coordination number $z$ in the rhombus tiling. Each panel shows a different interaction strength $U$. Different $z$ values are displayed in a light-blue to dark-purple color scale.
Both DQMC and analytical results for  $A_z(\omega=0)$ are shown in the non-interacting case in panel (a). A crossing between the curves is observed in panels (d)-(f), signaled by a vertical line. The insets in (b)-(f) present  $\Delta TW_{z}(T) \equiv TW_{z}(T) - TW_{7}(T)$ to further highlight the presence or absence of crossing points.
The range of crossing points between any pair $[TW_{z}(T), TW_{z'}(T)]$ is used to determine the error bars shown in Fig.~\ref{fig:model} (c).
}
\label{fig:proxy}
\end{figure}

The behaviors of  $S_{AF}$ and $C$ in the quasicrystal are very similar to what is seen in periodic lattices undergoing a Mott transition \cite{paiva05}. Meanwhile, the distinguishing feature of the quasicrystal is that its sites have multiple coordination numbers $z$, which contribute unevenly to the zero-energy peak (i.e. the ``flat band''). It is thus desirable to probe the $z$-resolved spectral function $A_z(\omega) = -\mathrm{Im} G_z(\omega)/\pi$, which encodes the LDOS for a site with coordination number $z$. Here, $G_z(\omega)$ is the local Green's function averaged over non-boundary sites with same $z$. While obtaining the frequency-dependent function requires analytical continuation, $A_z(\omega=0)$ can be estimated via the proxy:

\begin{align}
W_z(T) = \frac{\beta}{\pi} G_z(\tau=\beta/2) = \beta \int \frac{d\omega}{2\pi} \frac{A_z(\omega)}{\cosh (\beta \omega/2)} \,,
\end{align}
which involves only the imaginary-time Green’s function \cite{trivedi95,Mendl17,Wang20}. Indeed, as shown in Fig.~\ref{fig:proxy}(a) for the non-interacting case ($U=0$), the LDOS proxy $W_z(T)$ approaches the analytically calculated $A_z(\omega=0)$ as $T \rightarrow 0$. The fact that the proxy only approaches a non-zero value for $z=3$ and $z=5$, with $W_3(T \rightarrow  0) \gg W_5(T \rightarrow  0)$, is a consequence of the zero-energy peak being strongly dominated by the $z=3$ sites, as shown in Fig.~\ref{fig:model}(b).

As $U$ increases, two different behaviors emerge in the weak and strong coupling regimes, see Figs.~\ref{fig:proxy}(b)-(f). For $U=2$ [panel (b)], while all LDOS proxy curves $W_z(T)$ are suppressed with respect to their noninteracting values,$W_3(T)$ remains the largest one as $T\rightarrow0$, i.e. the low-energy DOS remains dominated by the $z=3$ sites. Meanwhile, for $U=8$ [panel (f)], the situation is different. Above a characteristic temperature $T^* = 1.85$ (vertical black line), all $W_z(T)$ curves are rather similar, with $W_3$ slightly larger than the other curves, and $W_7$ slightly smaller. At $T^*$, however, the $W_z(T)$ curves cross 
(highlighted by the inset), and the hierarchy of the LDOS proxy curves is reversed below $T^*$, with $W_3(T)$ becoming the smallest one. 
Thus, in the strong-coupling regime, the sites with the smallest coordination numbers are more strongly impacted by the interaction, in that their LDOS proxies are more suppressed than the LDOS proxies of the sites with the largest coordination numbers -- despite the fact that the former dominate the low-energy region of the non-interacting spectrum. A similar behavior is observed for $U=7$ and $U=6$ [panels (e) and (d)], with $T^*$ decreasing for decreasing $U$, while a clear crossing cannot be resolved for $U=5$ [panel (c)]. Similar behaviors are observed in the kite-and-dart tiling (see SM \cite{sm}).

It is illuminating to plot the crossing temperatures $T^*$ as a function of $U$, see Figs. ~\ref{fig:model}
(c) [rhombus] and (f)[kite-and-dart]. A linear interpolation of the non-zero $T^*$ points gives an extrapolated $U_\mathrm{ext}$ value, defined as $T^*(U_\mathrm{ext})\rightarrow0$, that is close to the critical value $U^*$ associated with the Mott transition obtained from the specific heat curves. While it is not clear why or even whether the relationship $T^*(U)\sim (U-U^*)$ holds, this simple analysis lends further support to the conclusion that the interaction impacts different sites in distinct ways inside the Mott phase.

\begin{figure}
\centering
\includegraphics[width=1\columnwidth]{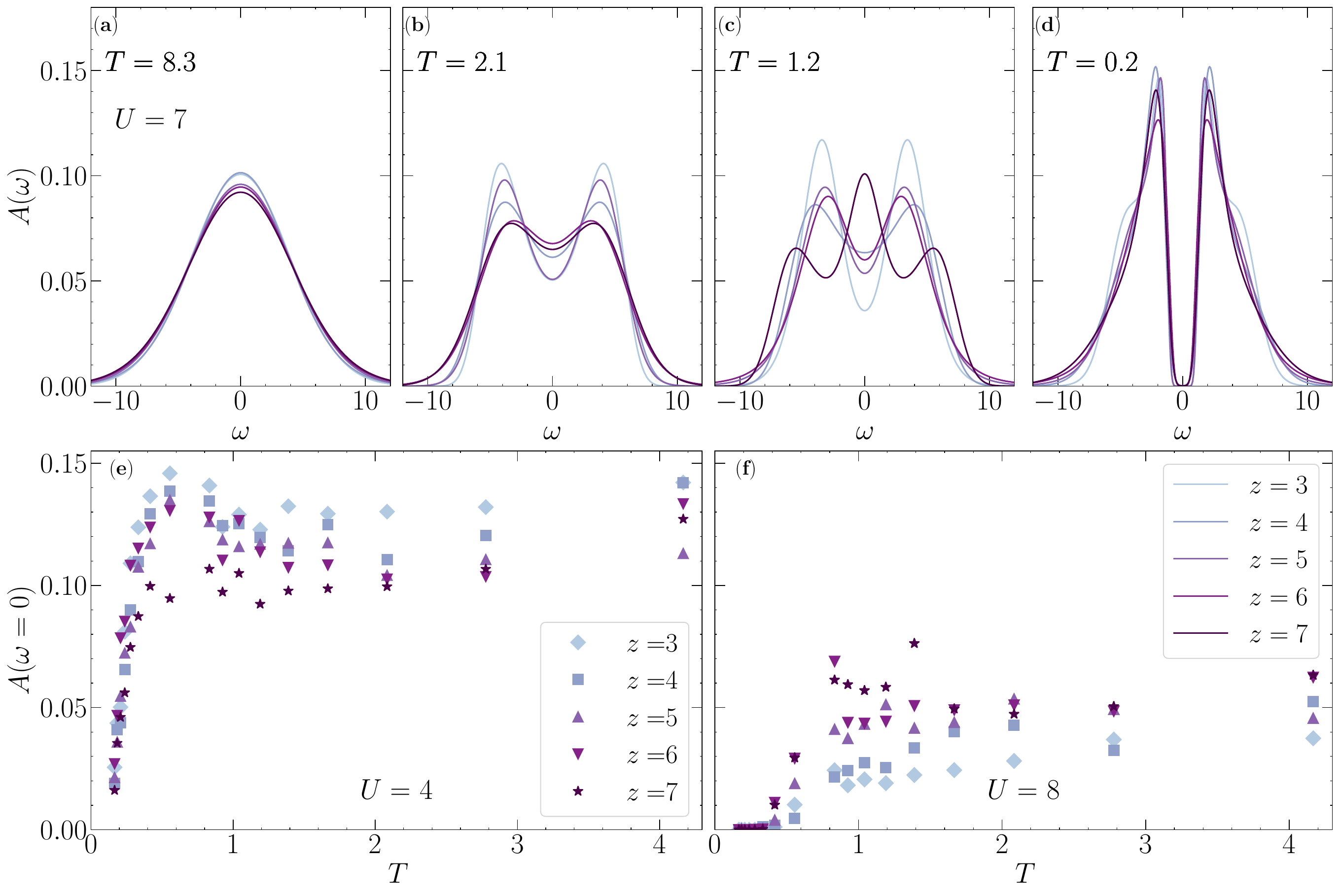}
\caption{(a)-(d) Evolution of the $z$-projected spectral function $A_{z}(\omega)$, obtained via analytical continuation, as temperature $T$ is lowered for fixed $U=7$ for the rhombus lattice.
(e)-(f) Zero-energy spectral function $A_{z}(0)$ as a function of $T$ for $U=4$ and $U=8$. In (e), $A_3(0)$ is generally the largest wheres in (f), it is generally the smallest. A gap is clearly seen for all $z$ values in (f). 
}
\label{fig:Aw}
\end{figure}

To further validate our findings of site-selective correlations, we perform an analytical continuation of $G_z(\tau)$  \cite{kaufmann23} to extract the $z$-resolved spectral functions  $A_z(\omega)$. Figs.~\ref{fig:Aw} (a)-(d) show their temperature dependence for $U=7$. At high temperatures, $A_z(\omega)$ has the typical metallic shape, with $A_3(\omega=0)$ being the largest. However, as temperature is lowered and a gap starts to form, the site-selective character of the transition emerges. First, the separation between the finite-$\omega$ peaks is larger for $A_3(\omega)$. Second, as the temperature is lowered even further, a Kondo-resonance peak at $\omega=0$ can be seen in $A_7(\omega )$. Interestingly, such a peak often emerges in single-site DMFT simulations of the Hubbard model \cite{Jarrell1993,Kotliar2000}. Since DMFT is exact in the limit $z\rightarrow \infty$, it is reasonable that mean-field features in $A_z(\omega)$ emerge for larger $z$. 

To perform a more quantitative analysis, Figs. ~\ref{fig:Aw} (e) and (f) show the $T$ dependence of the spectral function evaluated at zero energy, $A_z(\omega=0)$, for $U=4$ and $U=8$, respectively. For $U=4$, the hierarchy observed from the curves is that $A_3(0)$ is the largest for all temperatures. Conversely, for $U=8$, the hierarchy is reversed at low temperatures, with $A_3(0)$ generally assuming the smallest value. Moreover, a full gap is observed at low enough temperatures.

Our result of site-selective correlations in interacting quasicrystals qualitatively agrees with a RDMFT investigation of the doped Hubbard model on the rhombus tiling  \cite{sakai22}, which found that added carriers dope more the sites with smaller (larger) $z$ at weak (strong) coupling. Our finding is also reminiscent of the site-selective behavior observed in a recent CPA study of the Hubbard model on commensurately-twisted tetragonal bilayers  \cite{Jiang2024}. More broadly, the concept that correlations affect different sites in distinct ways generalizes the notion of orbital-selective correlations, by which different orbitals experience the effects of correlations differently \cite{DeMedici05,Vojta2010}. In analogy to the scenario of an orbital-selective Mott transition, it is an interesting question whether a site-selective Mott transition could emerge in quasicrystals.

In summary, our DQMC simulations demonstrate that, in interacting quasicrystals, the onsite repulsion affects sites with different coordination numbers in distinct ways. Specifically, the LDOS is more strongly suppressed on sites with the smallest $z$, which for the two types of Penrose quasicrystals studied here correspond to $z=3$. Interestingly, in a mean-field approach to the Hubbard model on a periodic lattice, the critical $U^*$ signaling the Mott transition scales as $U^* \sim 1/z$, since the bandwidth scales as $z$ \cite{Thomas2017}. It is intriguing that in a quasicrystal, where $z$ is defined locally and there is no bandwidth, the sites with smallest $z$ are more strongly affected by interactions. More importantly, in the Penrose tilings investigated here, the $z=3$ sites are the ones that contribute the most to the zero-energy peak in the global DOS, which play the same role as a flat band in periodic lattices. Thus, our results reveal a subtle interplay between flat-band physics and interactions, pointing to the importance of the real-space structure of flat bands. Indeed, while featureless in momentum space, flat-bands often display non-trivial patterns in real space. This is not exclusive to quasicrystals: in the kagome lattice, the flat band has a non-trivial projection onto the sublattices \cite{Zeng2024}. Similarly, in twisted bilayer graphene, the flat band emerges primarily from the AA stacking sites \cite{Bernevig2022}. It will be interesting to investigate whether site selectivity also emerges in these flat-band settings.

\begin{acknowledgments}
We thank J. Schmalian for useful discussions. Y.Z. and R.M.F were supported by the Air Force Office of Scientific Research under Award No. FA9550-21-1-0423. R.S. was supported by the grant DE-SC0014671 funded by the U.S. Department of Energy, Office of Science. 
\end{acknowledgments}

\bibliography{quasiHub}

\end{document}


\title{Supplemental Material for \\
\textit{Site-selective correlations in interacting ``flat-band'' quasicrystals"}}

\author{Yuxi Zhang}
\affiliation{Department of Physics, University of Illinois Urbana-Champaign, Urbana, IL, USA}
\author{Richard T. Scalettar}
\affiliation{Department of Physics, University of California, Davis, CA, USA}
\author{Rafael M. Fernandes}
\affiliation{Department of Physics, University of Illinois Urbana-Champaign, Urbana, IL, USA}
\date{\today}
\maketitle

\vskip0.10in

\section{Details of the determinantal quantum Monte Carlo approach} \label{DQMC}

We employed the standard determinantal quantum Monte Carlo (DQMC) method in this work\cite{Hirsch1981}.
After discretizing the inverse temperature $\beta$ in the imaginary time axis $\beta=L_{\tau}\Delta \tau$, one can decouple the kinetic energy $K$ and the interaction $V$ of the Hamiltonian, since the penalty can be minimized by setting $\Delta \tau \rightarrow 0$.
After decoupling, the Hubbard interaction term can be dealt with via the (discrete) Hubbard-Stratonovich (HS) transformation
\begin{align} 
    e^{-U \Delta \tau (n_{\mathbf{i} \uparrow}-\frac{1}{2}) (n_{\mathbf{i}\downarrow}-\frac{1}{2})} = \frac{1}{2} e^{-\frac{U\Delta \tau}{4}} \sum_{S_\mathbf{i}} e^{\lambda S_\mathbf{i} (n_{\mathbf{i} \uparrow}-n_{\mathbf{i} \downarrow})} \,,
\end{align}
where $\cosh{\lambda} = e^{\frac{U\Delta \tau}{2}}$, and the additional HS field $S_\mathbf{i} = \pm 1$ is an Ising-like variable.
The partition function can be written as
\begin{align} \label{chap2eq:pathintegral}
    Z & = \sum  \Tr \, e^{-\beta \hat H} = \sum  \mathrm{Tr} \,\prod e^{-\Delta \tau \hat H}  \nonumber \\
      & = \sum  \Tr \,\prod e^{-\Delta \tau \hat K} e^{-\Delta \tau \hat V} \nonumber \\
      & = \sum  \det [\mathbf{I} + \prod e^{-\Delta \tau \mathbf{K}} e^{-\Delta \tau \mathbf{V}(S_{\mathbf{i},l})} ] \\
      & = \sum  \det \mathbf{M}\,,
\end{align}
where the sum is over the HS field $\{S_{\mathbf{i},l}\}$ and the trace is over the fermion operator.

Measurements are carried out via Green's functions and Wick's construction
\begin{align}
    G_{\sigma}(i,j) = \langle \hat{c}^{\phantom{\dagger}}_{\mathbf{i}\sigma} \hat{c}^{\dagger}_{\mathbf{j}\sigma} \rangle =
    [\mathbf{M}_{\sigma}]^{-1}_{ij} \,.
\end{align}
For example, the anti-ferromagnetic (AF) structure factor is defined as~\cite{Khatami2014} 
\begin{align}
  S_{AF} = \sum_{\mathbf{r}} (-1)^{\mathbf{r}} m(\mathbf{r}) \,,
\end{align}
where the spin-spin correlation functions are defined as
\begin{align}
m(\mathbf{r}) = \frac{1}{3} [2 m^{xx}(\mathbf{r})+m^{zz}(\mathbf{r})]
\end{align}
and
\begin{align}
m^{xx}(\mathbf{r}) = \langle S^{-} (\mathbf{i}+\mathbf{r}) S^{+} (\mathbf{i})\rangle \nonumber \\
m^{zz}(\mathbf{r}) = \langle S^{z} (\mathbf{i}+\mathbf{r}) S^{z} (\mathbf{i})\rangle \,,
\end{align}
where
\begin{align}
S^{+} (\mathbf{i}) & = c^{\phantom{\dagger}}_{\mathbf{i} \downarrow} c^{\dagger}_{\mathbf{i} \uparrow} \nonumber \\
S^{-} (\mathbf{i}) & = c^{\phantom{\dagger}}_{\mathbf{i} \uparrow} c^{\dagger}_{\mathbf{i} \downarrow} \nonumber \\
S^{z} (\mathbf{i}) & = \frac{1}{2} (n_{\mathbf{i} \uparrow}-n_{\mathbf{i} \downarrow})
\end{align}
can be directly accessed via Green's function.

To ensure reliable statistical results, we used $\Delta \tau = 0.1$, as well as Monte Carlo steps of $N_\mathrm{warms} =5,000$ - $10,000$ for thermalization sweeps and $N_\mathrm{meas}=80,000$ for measurement sweeps.

\section{Results for the kite-and-dart tiling} \label{kad}

While the main text focused on the results for the rhombus tiling, similar results were also obtained for the kite-and-dart tiling. Here, we present the analysis of the DQMC results for the Hubbard model on the kite-and-dart tiling, plotting the same quantities as those shown in Figs. 2-4 of the main text for the rhombus tiling. 

\begin{figure*}[]
\centering
\includegraphics[width=1\columnwidth]{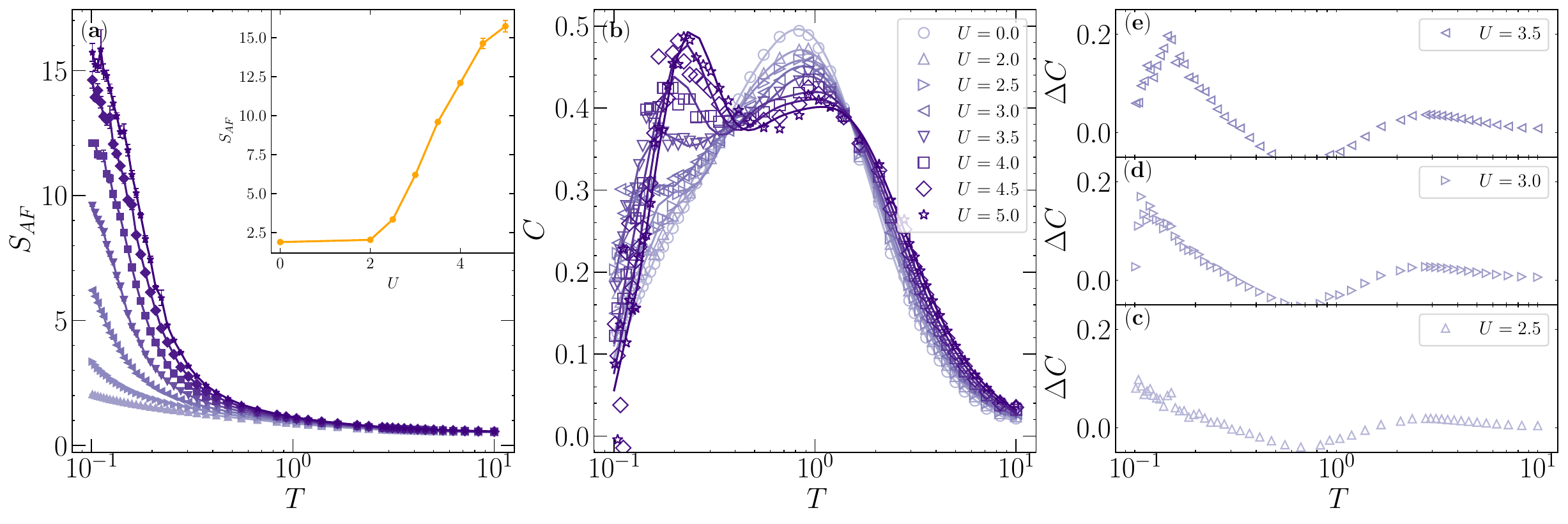}
\caption{Anti-ferromagnetic phase transition on the kite-and-dart tiling.  Panel (a): AF structure factor $S_{AF}$; panel(b): specific heat $C$, both as a function of temperature $T$ on a logarithmic scale.
Increasing interaction strength $U$ is distinguished by the color scale from light-purple to dark-purple. $S_{AF}$ behavior at low temperature $T=0.1$ is highlighted in panel (a) inset.
Panels (c)-(e) show $\Delta C$, defined as $C(U)-C(U=0)$, for three $U$ values: $U=2.5$, $U=3.0$ and $U=3.5$ (from bottom to top).}
\label{fig:mott}
\end{figure*}

Fig. \ref{fig:mott} shows the anti-ferromagnetic structure factor $S_{AF}$  and specific heat $C(T)$ as a function of temperature for various interaction strength values. Similarly to the rhombus case, AF order develops when temperature is lowered for sufficiently strong coupling strength.
The behavior of specific heat $C(T)$ shows the same trend as the rhombus case: for small interaction values, a broad one-peak shape signals a non-ordered phase, whereas for $U \geqslant 3.0$, a two-peak structure signals the onset of the ordered magnetic state. Thus, we estimate $U_c = 3.0$, slightly smaller than the value obtained for the rhombus tilinge.

\begin{figure}[htbp!]
\centering
\includegraphics[width=0.7\columnwidth]{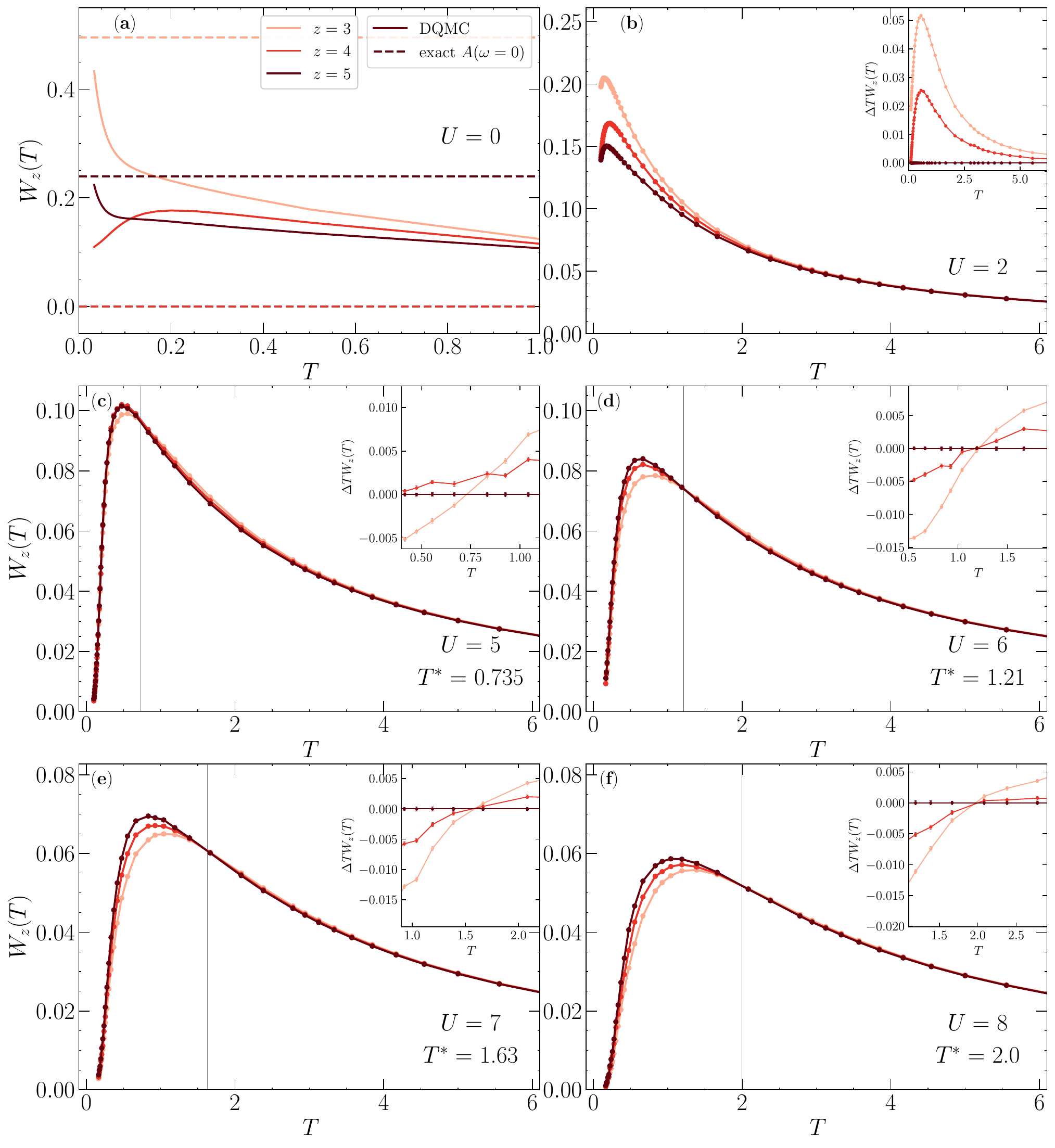}
\caption{The $z$-projected zero-energy spectral function proxy $W_{z}(T)$ (see main text for definition) with increasing interaction strength, in the case of the kite-and-dart tiling. The insets in panels (b)-(f) show $\Delta TW_{z}(T) \equiv TW_{z}(T)-TW_{5}(T)$, highlighting the crossing points marked by the vertical line in the main panels.
}
\label{fig:proxy}
\end{figure}

Like in the rhombus case, a site-selective behavior for the correlation effects is seen on the kite-and-dart tiling. Fig. \ref{fig:proxy} shows the proxy of the $z$-projected zero-energy spectral function $W_{z}(T)$ for the kite-and-dart case. In the weak coupling case $U=2$, displayed in panel (b), the sites with the smallest coordination number $z=3$ have the largest spectral function proxy; these are also the sites that contribute the most to the zero-energy peak of the non-interacting density of states. 
In the strong coupling case $U \geqslant 5.0$, however, a reversal occurs and the sites with $z=3$ display the smallest spectral function proxy below a characteristic temperature  $T^{*}$ where the $W_z(T)$  curves for different $z$ values cross. 

\begin{figure}[htbp!]
\centering
\includegraphics[width=0.8\columnwidth]{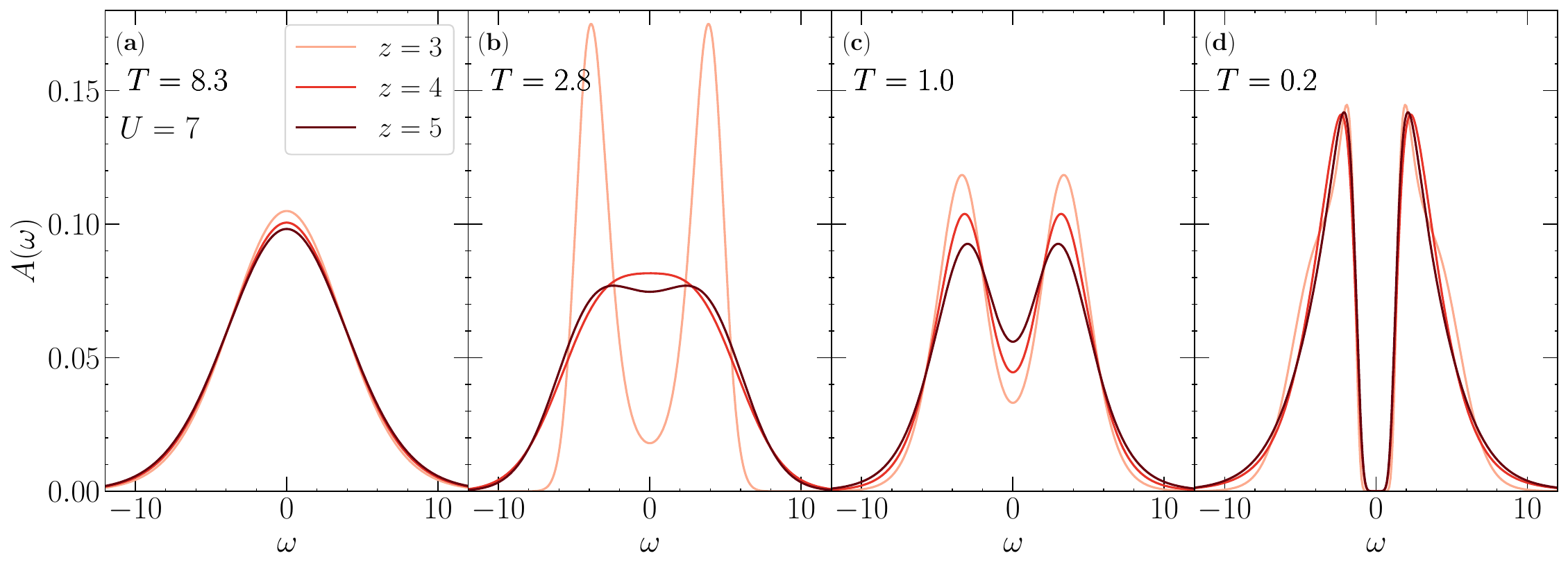}
\caption{The $z$-projected spectral function $A_{z}(\omega)$ obtained via analytical continuation for $U=7$ and at various temperature values $T$, in the case of the kite-and-dart tiling.
}
\label{fig:Aw}
\end{figure}

\begin{figure}[htbp!]
\centering
\includegraphics[width=0.8\columnwidth]{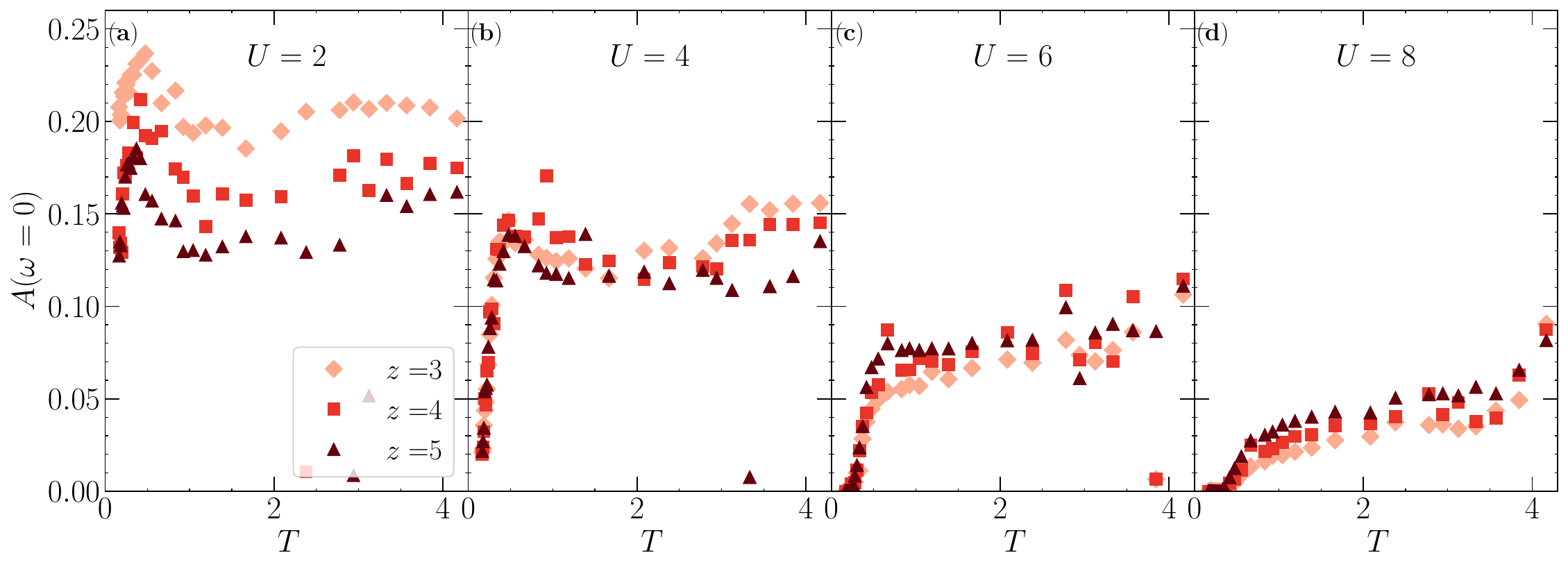}
\caption{The $z$-projected spectral function $A_{z}(\omega)$ obtained via analytical continuation for various fixed interaction strength values $U$, as a function of temperature $T$, in the kite-and-dart tiling.
}
\label{fig:Aw0}
\end{figure}

Finally, we perform an analytical continuation to obtain the $z$-resolved spectral function $A_z(w)$ for the kite-and-dart geometry.
Fig.~\ref{fig:Aw} shows  $A_z(w)$ at strong coupling $U=7$ as temperature is lowered;
On the one hand, a gap opens at the lowest temperature $T=0.208$, corresponding to the Mott phase as discussed in the main text. 
On the other hand, comparing panels (a) and (c), we see the reversal in the hierarchy of $A_z(\omega=0)$ values: at higher temperatures, $A_3(0) > A_4(0) > A_5(0)$, whereas at low temperatures, $A_5(0) > A_4(0) > A_3(0)$. A systematic analysis of the temperature and interaction-strength dependence of $A_z(0)$ is shown in Fig.~\ref{fig:Aw0}. For $U=2$, $A_3(0)$ is generally the largest whereas for $U=8$, $A_5(0)$ is generally the largest. It is interesting to note that for intermediate values of $U$, e.g. $U=4$, there are regimes where $A_4(0)$ is the largest. This could be a consequence of the fact that in the kite-and-dart tiling, the $z=5$ sites also contribute to the zero-energy peak in the global DOS, although the dominant contribution still comes from the $z=3$ sites. We note that, for
strong interactions $U \geqslant 6.0$, a gap opening can be seen approaching low temperatures, signaling the Mott transition.